\documentclass[10pt,twocolumn,twoside] {IEEEtran} 

\usepackage{multirow}
\usepackage{graphicx}
\usepackage{epstopdf}
\usepackage{latexsym}
\usepackage[tight,footnotesize]{subfigure}
\usepackage{multirow}
\usepackage{longtable} 
\usepackage{algorithm}
\usepackage{algpseudocode}

\usepackage{cite}

\usepackage{cite}

\usepackage[cmex10]{amsmath}
\usepackage[tight,footnotesize]{subfigure}

\title{Human Computer Interaction Using Marker Based Hand Gesture Recognition}
\author{\IEEEauthorblockN{Sayem Mohammad Siam\IEEEauthorrefmark{1}, Jahidul Adnan Sakel, and Md. Hasanul Kabir\\}
\IEEEauthorblockA{Department of Computer Science and Engineering, Islamic University of Technology,\\
Gazipur--1704, Bangladesh \\
\IEEEauthorrefmark{1}sayemsiam@gmail.com}
}

\begin{document}
%
\maketitle

\begin{abstract}
\pagenumbering{roman} \setcounter{page}{1} 
Human Computer Interaction (HCI) has been redefined in this era. People want to interact with their devices in such a way that has physical significance in the real world, in other words, they want ergonomic input devices. In this paper, we propose a new method of interaction with computing devices having a consumer grade camera, that uses two colored markers (red and green) worn on tips of the fingers to generate desired hand gestures, and for marker detection and tracking we used template matching with kalman filter. We have implemented all the usual system commands, i.e., cursor movement, right click, left click, double click, going forward and backward, zoom in and out through different hand gestures.
Our system can easily recognize these gestures and give corresponding system commands. Our system is suitable for both desktop devices and devices where touch screen is not feasible like large screens or projected screens.
\end{abstract}
\section{Introduction}

The field of Human Computer Interaction (HCI) aims to improve interactions between users and computers by making computers more usable and receptive to the user's need over the last few years. This field has developed many input-output techniques including the technique using hand gestures as input signals. We propose a hand gesture recognition technique that uses colored markers. The system inputs, traditionally given by a mouse or touch-pad, can be performed using our proposed system. That is, the user will be able to perform system commands with the different hand gestures.\\	

To develop a cost effective system that allows user to interact with devices using hand gestures with the help of colored markers, we have to overcome several challenges such as detection of the markers in different light intensities, real time detection and tracking, accurate gesture recognition, segmenting out the marker from the same colored background.
We have worked with hue and saturation values of the image so the effect of light intensity has been reduced during detection \cite{hsi1}.
The use of template matching \cite{templateMatching2,jong,vina} using kalman filter\cite{kalman1,kalman2,kalman3} to detect and track the color marker has reduced the computational complexity of usual template matching. 
To separate the marker from the same color background object, we define a range of acceptable area for the marker size, if the area is greater than or smaller than the defined range then that object is considered as part of the background rather than a marker. Another challenge is the tracking of the marker when some other object having same color and size as the marker comes in the image. This situation has been managed by the use of kalman filter\cite{kalman1,kalman2,kalman3}. The next challenge is accurate gesture recognition, which mostly depends on accuracy of detection and tracking. Last but not the least, is the challenge to make the whole system user friendly. The solution to this lies mostly in the solution of the other aforementioned challenges.\\

The paper is organized as follows. In section \ref{Related Work}, we discuss about existing human computer interaction techniques with some detection and tracking techniques.
We discuss our proposed method in details including detection and tracking algorithms and the proposed hand gestures with their recognition techniques in section \ref{Proposed Method}. Section \ref{Experimental Analysis} shows user friendliness of our system, gesture generation accuracy and average performance analysis. Finally,
conclusions are drawn and future plans are discussed in section \ref{Conclusion}.
\section{Related Work}
\label{Related Work}
Gesture recognition is mainly divided into two types, “Data-Glove based” and “Vision based” approaches. The Data-Glove based methods use sensor devices for capturing hand and finger motion data as input. The extra sensors make it easy to recognize exact hand gesture. However, the extra devices are quite cumbersome and expensive \cite{mulder}. In contrast, the Vision based methods\cite{quek} require only a camera without the use of any extra devices. Several systems have been developed for interating with computers via hand gesture or gesture of other body parts of human. Pranav Mistry, et al. \cite{mit_sixth_sense,mistry2,mistry3} proposed a method named wearable gesture interface which is not only an input but also an output tool.
Kinect\cite{kinect2,kinect} is a motion sensing input device by Microsoft for the Xbox 360 video game console and Windows PCs. Based around a webcam-style add-on peripheral for the Xbox 360 console, it enables users to control and interact with the Xbox 360 without the need to touch a game controller, through a natural user interface using gestures and spoken commands.
M. Baldauf and P. Fröhlich\cite{matthias} present a framework for spotting hand gestures that is based on a mobile
phone, its built-in camera and an attached mobile projector as medium for visual feedback.
J. Gips et al.\cite{Betke,camera_mouse2} developed a system that uses a camera to visually track the tip of the nose or the tip of a finger or some other selected feature of the body and moves the mouse pointer on the screen accordingly.
W. Hürst and C. V. Wezel\cite{wolfgang} proposed new interaction metaphors for augmented reality on mobile phones, i.e., applications where users look at the live image of the device's video camera and 3D virtual objects enrich the scene that they see.
Some of those use cam shift method, which is based on histogram back projection\cite{Swain} technique and others use cross-correlation\cite{james} for tracking. For detection different types of techniques are used, Feature-Based Matching and Template matching are most popularly used \cite{robert}. Normalized correlation co-efficient has been used in camera mouse \cite{Betke,camera_mouse2}. Chawalitsittikul and Suvonvorn \cite{Chawalitsittikul} used black hand gloves with colored markers and Cam-Shift technique for building their system. Wang and Robert \cite{wang,robert} proposed complex patterns of glove with specific tracking methods in order to retrieve the 3D hand pose for manipulating object in three dimension. Sýkora, et al. \cite{daniel} focused on how to track the color ball for augmented reality application.
\section{Proposed Method}
\label{Proposed Method}
We propose a method that uses the marker detection and tracking technique. For detection and tracking we use template matching algorithm\cite{templateMatching2,jong,vina} with kalman filter\cite{kalman1,kalman2,kalman3} since our marker size and color is fixed. The use of colored marker has increased the accuracy in detection and reduced the computation complexity which is very crucial for real time applications. We have used only two colored markers namely red and green to generate eight hand gestures to give commands to the desktop or laptop computer, that has a consumer grade camera, instead of using mouse or touch-pad. The integration of sliding window mechanism\cite{haiqian} with the template matching has reduced its time complexity. For greater accuracy of detection and smooth movement of system cursor we have used kalman filter\cite{kalman1,kalman2,kalman3}. We have worked with the hue and saturation values of HSI color model with a view to reduce the effect of different light intensity on the image \cite{Manresa,hsi1,book1,Gonzalez}.

\subsection{System Overview}
\begin {figure}[]
\centering 
\includegraphics[height=231px,width=120px]{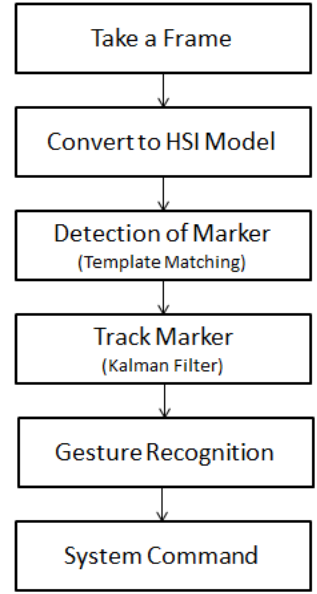}
\caption{System analysis diagram of proposed method}
\label{systemoverview} 
\end {figure}
The figure \ref{systemoverview} shows the system overview of our proposed method. First the web-cam takes a frame and converts it to HSI model, that is we use the hue and saturation information of the image. Then the marker detection is done by the template matching as we will discuss in section \ref{matching}. The tracking is done with the help of Kalman filter and will discuss in section \ref{kalmanFilter}. The gesture recognition is done as the system analysis diagramm illustrated in figure \ref{gesturerecognition}. As soon as the gesture is recognized, specific command is performed.  

\subsection{Choosing The Color Model}
Our most important challenge is that our system should have to work in different environments with different light intensities. If we use the RGB model we will not be able to detect our marker in different light intensities. To work in different light intensities we took HSI model because H and S value are light invariant\cite{hsi1}. We first get the R,G and B values from the input image then we converted it in H (Hue), S (saturation) and I (intensity)\cite{Gonzalez} values.

\subsection{Matching and Traversing Techniques}
\label{matching}
There are different type of matching techniques for marker recognition. Since the markers have fixed color and shape, we can use template matching for detection. There are different types of template matching \cite{lam}. For matching we have used SSD\cite{SSDNCC,SSDSAD} with sliding window mechanism\cite{haiqian} which has a very small time complexity. We modified the equation of general SSD to use both Hue and Saturation values of the HSI model. For a particular pixel in a input image we will get one Hue and one Saturation value and we can find out the response value using the following equation \ref{ssd}. When a pixel is actually center of the marker then its response value will be the least otherwise it will give higher value, i.e., matching pixel will give lower response value and non matching pixel will give higher response value. Let's assume our input image is M*N and our mask is m*n matrix. Here, the matrix horizontal axis is y and vertical axis x.
\begin{align}
\label{ssd}
RV_{x,y}=\sum_{s=-a}^{a}{\sum_{t=-b}^{b}{{\substack {w1*(H(x+s,y+b)-h)^2\\
+w2*(S(x+s,y+b)-s)^2}}} }
\end{align}
 Where $RV_{x,y}$ is the response value for the point x,y in the input image, mask size is m*n.\\ $a=(m-1)/2$ and $b=(n-1)/2$ and w1 and w2 are weight co-efficients. In our implementation we use $w1=w2=1$.

\subsection{Sliding Window Mechanism}
We are using using sliding window mechanism\cite{haiqian} for calculating the response value which reduces the computation time for SSD (Sum of Squared Differences)\cite{SSDNCC,SSDSAD}  technique. Using the equation \ref{ssd} we can generally calculate the response value. Below we have discussed how we calculate the response value using sliding window mechanism if our left pixel response value is known.
\begin{itemize}
\item Left to Right\\ \label{sec:left}
Let's assume, we know the response value ($RV_{x,y-1}$) of the green pixel in figure \ref{fig:slide1} and  all the pixels under the gray window would contribute in calculating the response value. Now for calculating the response value ($RV_{x,y}$) of green pixel in figure \ref{fig:slide2}, we have to calculate the pixels under this new gray window. Under this new gray window in figure \ref{fig:slide2}, only the rightmost column is new compared to the gray window in figure \ref{fig:slide1} and the darker regions are common in both figures \ref{fig:slide1} and \ref{fig:slide2} so if we only add the rightmost column values and subtract the leftmost column values from the previous response value ($RV_{x,y-1}$) we will get the new ($RV_{x,y}$). We can easily use the equation \ref{eq:left} to calculate the response value when the window slides from left to right.
\begin{figure}[]
\centering
\subfigure[]{
	\includegraphics[height=124px,width=150px]{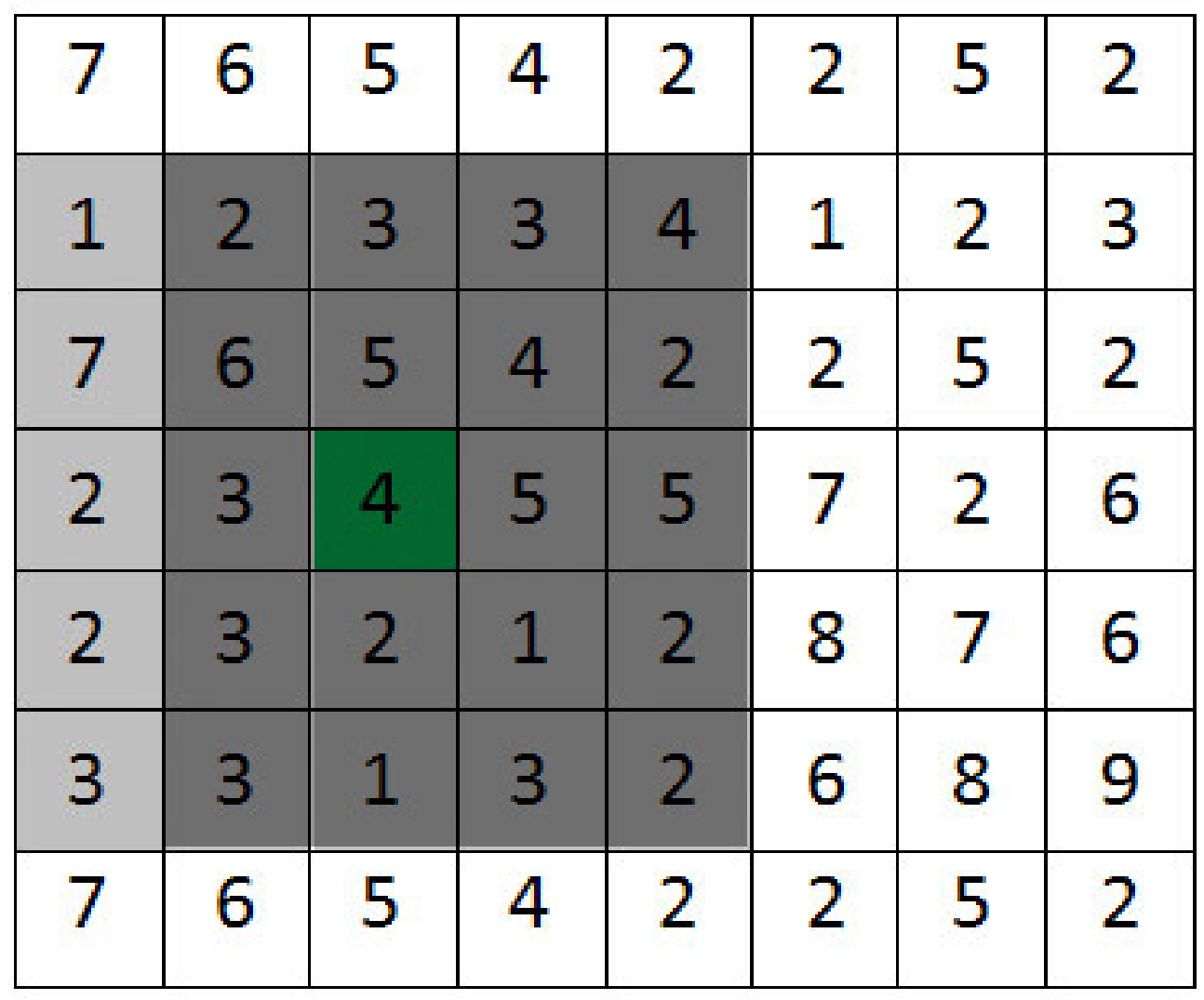}
	\label{fig:slide1}
}
\subfigure[]{
	\includegraphics[height=124px,width=150px]{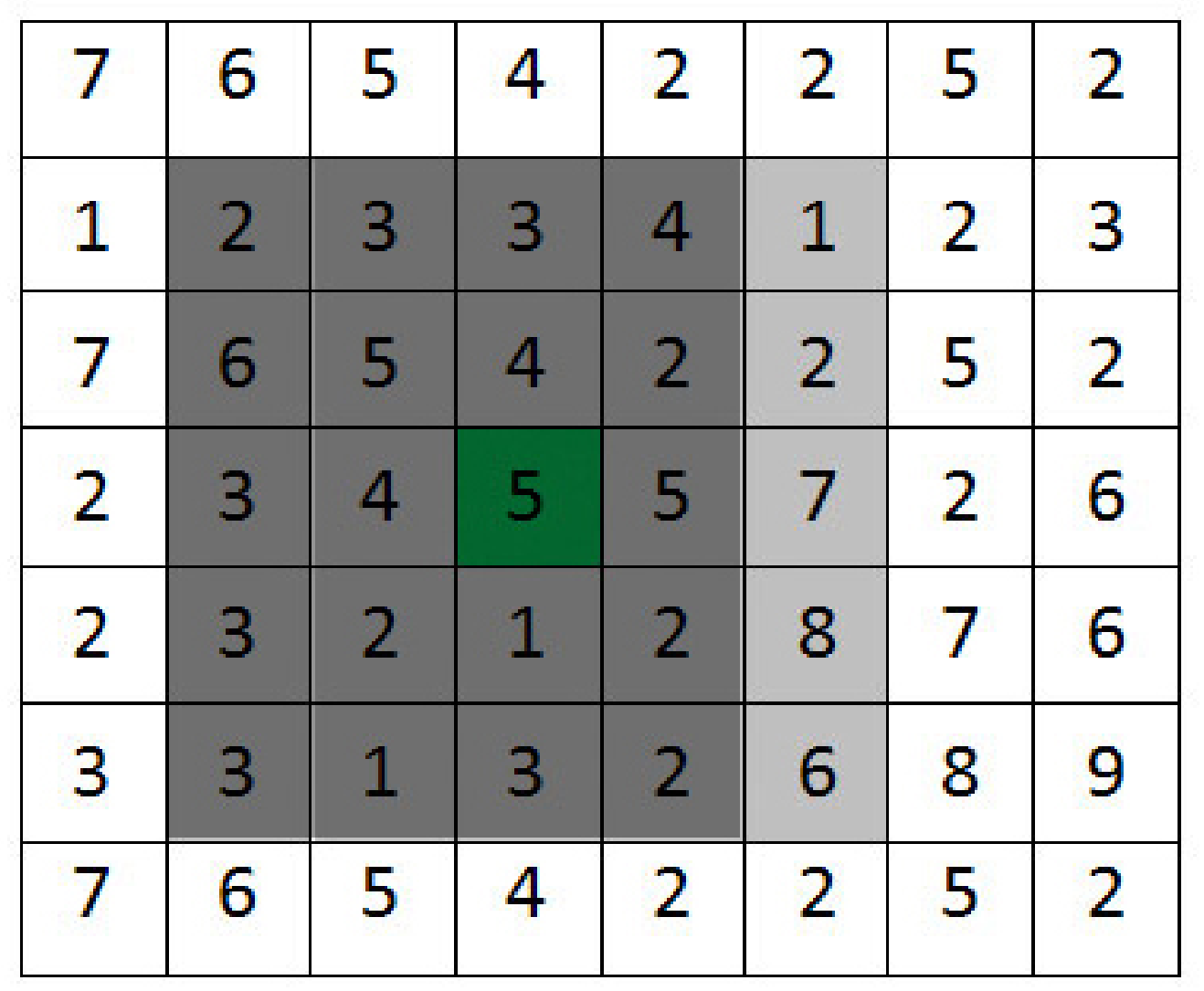}
	\label{fig:slide2}
}
\caption{Slided from left to right}
\label{fig:slide}

\end{figure}
\begin {figure*}[!t]
\centering 
\includegraphics[height=293px,width=450px]{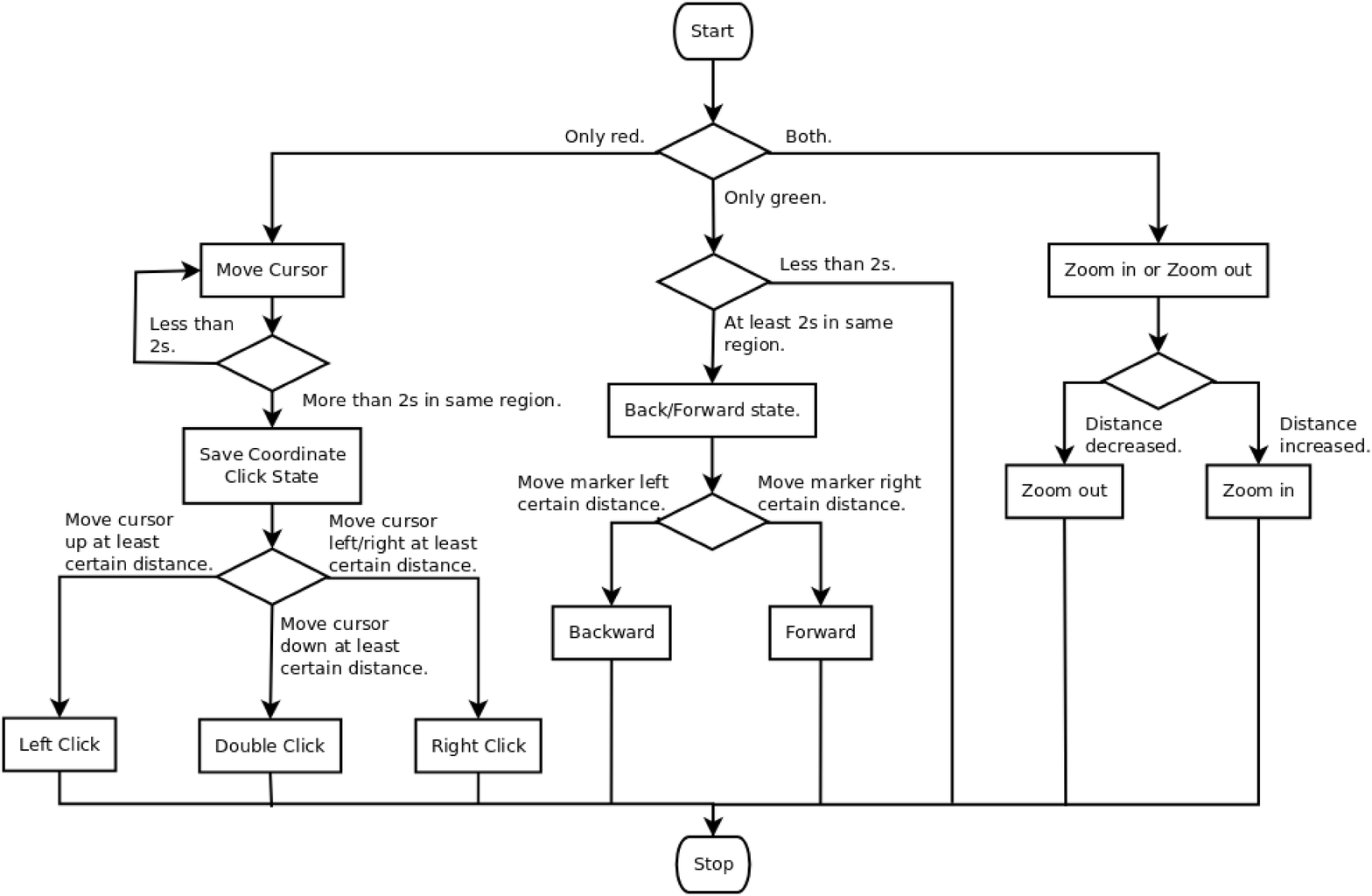}
\caption{System analysis diagram for gesture recognition}
\label{gesturerecognition} 
\end {figure*}
\begin{align}\label{eq:left}
add &=\sum_{s=-a}^{a}{\sum_{t=1}^{n}{\{\substack{ w1*(H(x+s,y+b+1-t)-h)^2\\+w2*(S(x+s,y+b+1-t)-s)^2\} }}} \nonumber\\
sub &=\sum_{s=-a}^{a}{\sum_{t=1}^{n}{\{\substack{ w1*(H(x+s,y-b-t)-h)^2\\+w2*(S(x+s,y-b-t)-s)^2\}}}}\nonumber\\
RV_{x,y}&=RV_{x,y-n}+add-sub \nonumber\\
\end{align}

\end{itemize}
In the same way we have used sliding window mechanism for top to bottom, right to left, bottom to up sliding to calculate the response value.
\subsection{Scanning Techniques}
For gesture recognition, we need to detect marker in each continuous frame. We scan each frame to detect the marker and raster or circular scanning technique is applied in each continuous frame, which are discussed in more detail below.
\begin{itemize}
\item Nth Distant Pixel Scanning:\\
For every frame, we are looking only at the nth distant pixels and calculating the response values.
\item Raster Scanning:\\
When the system starts for the first time or when the system loses the marker, it starts searching using raster scanning with nth distant pixel. It chooses a pixel as a center of the marker if the response value is lower than the threshold response value and meets the previously set marker size.
\item Circular Scanning:\\\label{circular}Circular scanning starts only if our system knows previous marker position. It starts from the previous point and circularly scan nth distant pixel and return the point if its response value is lower than the threshold response value and meets the previously set marker size. In circular scanning we define ``search window"\cite{Betke} that is centered at the position of the marker detected in the previous frame and within that search window marker will be searched for. If we do not find the marker in that search window it is assumed that marker has been lost and in the next frame we will search for marker by raster scanning.
\end{itemize}
\subsection{Center of Mass Calculation}
After detection of the marker we have to calculate the center of mass\cite{Chawalitsittikul}. If we do not detect the center of the marker, different pixels of the marker will be detected at different time and it will create wrong estimation of the marker position.

\subsection{Separating Marker From Background}
To meet the challenge of separation of the marker from the background first of all we choose markers as uniformly colored object. Secondly we have defined specific range of size of the marker so that it can be distinguished from a uniformly same colored background as the marker. The range is say [a,b] pixels, where a is the minimum area for the marker and b is the maximum area of the marker as in equation \ref{size}. If the marker area is found to be greater than b we simply consider that as background of marker's color and hence no marker is detected. Again if the marker area is found to be less than a we consider that as a red dot in the background and hence no marker is detected.
\begin{align}
\label{size}
a<S_{marker}<b
\end{align}
 where $S_{marker}$ is the size of the marker, $a$ is the minimum area for the marker and $b$ is the maximum area of the marker.
\subsection{Tracking Technique}
\label{kalmanFilter}
We are using Kalman filter \cite{kalman1,kalman2,kalman3} to track the marker smoothly and getting rid of unwanted jerks. The Kalman filter is a framework for estimating a process's state, and using measurements to correct or update these estimations. Our final goal is to move the cursor smoothly. If we only use the detection points to move the cursor, the cursor will not move smoothly because our marker movements will have some small sharp jerk due to some error in detection.
If we do not use Kalman filter we will not be able to place the cursor in a particular point for a second which is needed for our clicking purposes and most importantly we can not move the cursor in our desired way. If any point is misdetected as the center of marker and if we do not use Kalman filter the cursor will give a sudden jump for that misdetected point. Since the time and space complexity of kalman filter are not high, we can easily use this for our smooth detection purposes. Kalman filter is a recursive procedure, it only remembers the previous state and predicts the current state depending on the previous state and system models. Each point in a captured frame have been mapped with a point of the device's screen. This mapping, rather than depending solely on kalman filter, has made it easier for the user to have the cursor in desired location of the screen.

\subsection{Proposed hand gestures}

\subsubsection{Cursor Move}
The system cursor moves with the movement of red marker. If the position of the red marker is in upper-left corner of the captured frame then cursor will be at the upper-left corner of the screen. If the red marker moves left the cursor would move left and same for the right.

\subsubsection{Left Click}
Only red marker is involved. The red marker has to be within a certain small region (actually user would try to keep the marker still) for some specified time, i.e., 2 seconds. If the red marker is then moved upward it is considered as left click.

\subsubsection{Right Click}
Same as the ``left click" action except marker is to be moved right instead of moving upward.

\subsubsection{Double Click}
Same as the ``left click" action except marker is to be moved downward instead of moving upward.

\subsubsection{Zoom In}
Both the green and red markers are involved. If the distance between the markers are increased it is considered as a zoom in command.

\subsubsection{Zoom Out}
Both the green and red markers are involved. If the distance between the markers are decreased it is considered as a zoom out command.

\subsubsection{Forward}
Only green marker is involved. The green marker has to be within a certain small region (actually user would try to keep the marker still) for some specified time, i.e., 2 seconds. If the green marker is then moved right it is considered as Forward command.

\subsubsection{Backward}
Same as the ``Forward" action except marker is to be moved left instead of moving right.

\begin {figure}[]
\centering 
\includegraphics[height=110px,width=250px]{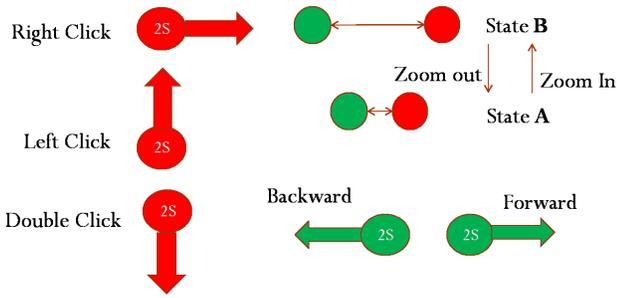}
\caption{Proposed hand gestures}
\label{handgesture} 
\end {figure}

\subsection{Gesture recognition}

The figure \ref{gesturerecognition} shows our logic to distinguish each gesture. Initially the system is at start state. The system looks for both red and green markers. If only red marker is detected the system goes to cursor move state. At this state if the system finds that the red marker is found within a certain region for a specified amount of time it saves the center of that region. Now if the cursor moves right the system goes to right click state (performs right click), else if the cursor moves upward the system moves to left click state (performs left click), else if the cursor moves downward the system moves to double click state (performs double click). If the system finds only the green marker while it is at start state the system checks whether the green marker is found within a certain region for a specified amount of time. If found the the system goes to Back/Forward state. Now if the green marker moves left the system moves to Backward state (performs backward command). Else if the green marker moves right the system moves to Forward state (performs forward command). At starting state if the system finds both red and green marker the system moves to Zoom in or Zoom out state. If the distance between two markers are decreased the system moves to Zoom out state (performs zoom out command). Else if the distance between two markers are increased the system moves to Zoom in state (performs zoom in command).

\section{Experimental Analysis}
\label{Experimental Analysis}
\begin {figure}[!t]
\begin {center}
\includegraphics[height=188px, width = 250px]{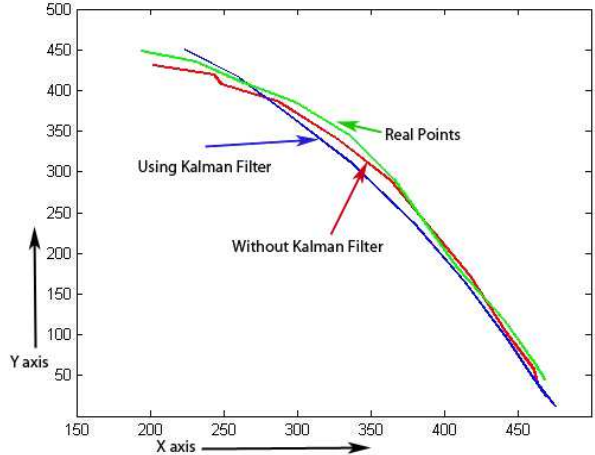}
\caption{Effect of using kalman filter}
\label{fig1}
\end {center}
\end {figure}
Our proposal requires a camera, red and green colored markers. We have implemented our system in Java. The laptop in which we carried our experiment had core-i5 processor and 4 GB of RAM. The web-cam was hp laptop web-cam. We did the experiments in different light intensity label: bright and dark environments.




\subsection{Impact of Using kalman Filter}


In figure \ref{fig1} the green line shows the actual position of the marker, the blue line shows the position of the system cursor and the red line shows the position of the system cursor if we did not use the Kalman filter. The red line has small undulations that shows some jerking that occurs in the absence of Kalman filter. On the other hand the blue line is quite smooth as is the green line. This smoothness makes the blue line more similar to the green line which results in good gesture generation.  

\subsection{Error Rate Between Experimental Detection and Actual Position}
\begin{table*}[tp]
\caption{Error rate between experimental detection and actual position} 
\centering  
\begin{tabular}{| c | c | c | c | c | c | c |} 
\hline\hline                        
frame no &  experimental X & experimental Y & Real X & Real Y & Error(pixel) & $\sigma$ of error\\ [0.5ex] 
\hline                  
1 & 476 & 13  & 469 & 46 & 33.73 &\\ 
2 & 470 & 25  & 462 & 65 & 40.79 &\\
3 & 261 & 416  & 231 & 436 & 36.05 & 8.27\\
4 & 224 & 451  & 194 & 449 & 30.06 &\\
5 & 295 & 310  & 320 & 350 & 32.02 & \\
\hline
\end{tabular}
\label{error in detection} 
\end{table*}

In table \ref{error in detection}, the 2nd and 3rd columns show the position of the center of the marker calculated by our proposed technique and the 4th and 5th columns show the real position of the center of the marker. The average distance from the calculated position and the real position of the marker is 34 pixels approximately. 6th column of table \ref{error in detection} shows that the standard deviation of the error is 8.27 which is quite small, this shows the distance between the actual position and the calculated position of the marker remains quite constant and hence cannot create problem in generation of gestures.   

\subsection{Losing Marker Vs Velocity}		
\begin{table*}[tp]
\caption{Losing Marker vs velocity} 
\centering  
\begin{tabular}{| c | c | c | c |} 
\hline\hline                        
time(s) & X(pixel) & Y(pixel) & Velocity(pixel/s)\\ [0.5ex] 
\hline                  
4.915 & 304 & 277 & \\ 
5.036 & 220 & 236 & \raisebox{1.5ex}{772.495} \\
 [1ex]      
\hline
5.506 & 117 & 264 & \\ 
5.752 & 359 & 248 & \raisebox{1.5ex}{985.888} \\
 [1ex]      
\hline
8.523 & 195 & 212 & \\ 
8.744 & 198 & 450 & \raisebox{1.5ex}{1077.01} \\
 [1ex]      
\hline
9.810 & 304 & 277 & \\ 
10.026 & 220 & 450 & \raisebox{1.5ex}{890.347} \\
 [1ex]      
\hline
11.812 & 423 & 256 & \\ 
12.011 & 363 & 110 & \raisebox{1.5ex}{793.206} \\
 [1ex]  
 \hline
\end{tabular}
\label{velocity} 
\end{table*} 

Table \ref{velocity} shows the effect of velocity of the movement of the hand of the user on the detection of the marker. If we move hands too quickly then our camera captures an image where markers become blurred. That's why our system can not detect the marker in that particular frame, i.e., our system loses the marker. In each row the second sub-row shows the actual position of the lost marker at time $t_2$. The first sub-row shows the position of the detected marker at time $t_1$. The fourth column shows the velocity which is responsible for losing the marker. The average velocity responsible for losing the marker is  greater than or equal to 903.79 pixel/s. This is totally dependent on the quality of the camera.
\subsection{Benefits of Using Search Window}
\begin{table*}[tp]
\caption{Time required to detect after losing} 
\centering  
\begin{tabular}{| c | c | c |} 
\hline\hline                        
frame no &  without boundary box(s) & with boundary box(s) \\ [0.5ex] 
\hline                  
1 & 0.124 & 0.083 \\
2 & 0.097 & 0.096 \\
3 & 01    & 0.11 \\
4 & 0.095	& 0.08 \\
5 & 0.117	& 0.082 \\
6 & 0.089	& 0.085 \\
7 & 0.1	& 0.087 \\
8 & 0.127	& 0.081 \\
9 & 0.08 	& 0.09 \\
10 & 0.1	& 0.086 \\
\hline
Avg & 0.1029 & 0.0884 \\
\hline
\end{tabular}
\label{detection time} 
\end{table*}

The table \ref{detection time} shows the effect of using search window\cite{Betke} which has been discussed in the section \ref{circular}. The second column shows the time required to detect the lost marker when we do not use the search window for circular scanning. The third column shows the time required to detect the lost marker when we use the search window for circular scanning. We can easily observe that the improvement is almost 12\% if we use the boundary box in the circular scanning.  
\subsection{Gesture Recognition Accuracy}
Gesture recognition accuracy is the comparison between the gesture formed by our tracking technique and the actual gesture that user intended to generate. From figure \ref{fig1} we can observe that the green line and the blue line generates quite similar gesture. Accuracy is not essential rather gesture pattern is.


\subsection{User-Friendliness}
To analyze the user-friendliness we asked five users to use our system and perform the actions cursor move, left click, right click, zoom in, zoom out, Forward and backward. Each user was asked to perform each command 30 times. In the first 10 attempts the overall accuracy was 59.38\% as shows the table \ref{user performance1}. From the table \ref{user performance1} we can also see the average accuracy of each command which is at the bottom line of the table. We can also observe that the average accuracy for all the command are 64.2\% , 61.4\% , 55.7\% , 62.8\% , 52.8\% for the user 1, 2, 3, 4 and 5 respectively. Table \ref{user performance2} and \ref{user performance3} also illustrate similar informations as mentioned for table \ref{user performance1} for the next 10 attempts and next to next 10 attempts respectively. From the table \ref{user performance1}, \ref{user performance2} and \ref{user performance1} we can easily observe that average performance for each command increases as user tries the same command again and again. After 30th attempt the overall average accuracy is 79.4\%. It is observed that the system performance will increase as the user becomes more accustomed to it.
 
\begin{table*}[tp]
\caption{User performance in 1st attempt} 
\centering  
\begin{tabular}{| c | c | c | c | c | c | c | c | l |} 
\hline\hline                        
User & Cursor move & Left click & Right click & Zoom in & Zoom out & Forward & backward & Average\\ [0.5ex] 
\hline                  
1 & 10/10 & 5/10 & 7/10 & 6/10 & 6/10 & 6/10 & 5/10  & 64.2\%\\ 
2 & 10/10 & 6/10 & 5/10 & 6/10 & 7/10 & 5/10 & 6/10  & 61.4\%\\
3 & 9/10  & 5/10 & 6/10 & 7/10 & 5/10 & 4/10 & 5/10  & 55.7\%\\
4 & 10/10 & 5/10 & 6/10 & 5/10 & 5/10 & 6/10 & 7/10  & 62.8\%\\
5 & 9/10  & 4/10 & 5/10 & 5/10 & 4/10 & 5/10 & 5/10  & 52.8\%\\ [1ex]      
\hline 
avg & 96\% & 50\% & 58\% & 58\% & 54\% & 52\% & 56\% & 59.38\% \\
\hline
\end{tabular}
\label{user performance1} 
\end{table*}

\begin{table*}[tp]
\caption{User performance in 2nd attempt} 
\centering  
\begin{tabular}{| c | c | c | c | c | c | c | c | l |} 
\hline\hline                        
User & Cursor move & Left click & Right click & Zoom in & Zoom out & Forward & backward & Average\\ [0.5ex] 
\hline                  
1 & 10/10 & 6/10 & 8/10 & 7/10 & 7/10 & 7/10 & 7/10  & 75.7\%\\ 
2 & 10/10 & 7/10 & 6/10 & 6/10 & 7/10 & 6/10 & 7/10  & 70.0\%\\
3 & 10/10  & 6/10 & 6/10 & 7/10 & 6/10 & 5/10 & 5/10  & 64.2\%\\
4 & 10/10 & 6/10 & 6/10 & 6/10 & 7/10 & 7/10 & 7/10  & 70.0\%\\
5 & 9/10  & 5/10 & 6/10 & 5/10 & 5/10 & 6/10 & 6/10  & 60.0\%\\ [1ex]      
\hline 
avg & 98\% & 60\% & 64\% & 62\% & 64\% & 62\% & 64\% & 67.98\% \\
\hline
\end{tabular}
\label{user performance2} 
\end{table*}

\begin{table*}[tp]
\caption{User performance 3rd attempt} 
\centering  
\begin{tabular}{| c | c | c | c | c | c | c | c | l |} 
\hline\hline                        
User & Cursor move & Left click & Right click & Zoom in & Zoom out & Forward & backward & Average\\ [0.5ex] 
\hline                  
1 & 10/10 & 9/10 & 9/10 & 8/10 & 9/10 & 8/10 & 7/10  & 85.7\%\\ 
2 & 10/10 & 8/10 & 7/10 & 7/10 & 8/10 & 8/10 & 7/10  & 78.5\%\\
3 & 10/10  & 7/10 & 7/10 & 8/10 & 8/10 & 7/10 & 7/10  & 77.1\%\\
4 & 10/10 & 7/10 & 8/10 & 7/10 & 8/10 & 9/10 & 8/10  & 81.4\%\\
5 & 10/10  & 7/10 & 7/10 & 7/10 & 6/10 & 7/10 & 8/10  & 74.3\%\\ [1ex]      
\hline 
avg & 100\% & 76\% & 76\% & 74\% & 78\% & 78\% & 74\% & 79.4\% \\
\hline
\end{tabular}
\label{user performance3} 
\end{table*}
\section{Conclusion}
\label{Conclusion}
Our aim is to develop a cheap, hand-gesture-based input technique. In our proposed method we have used template matching which has been modified with the integration of the sliding window technique. We have modified scanning technique of usual template matching as well. Our proposed method processes each image in average at the rate of 1 frame per .025 second which makes it possible to interact with computer in real time. We are using markers which are just some pieces of color cloth and the web cam that comes with the laptop. So our system is very cheap. We have checked the user friendliness of our system which is quite satisfactory. So far we have worked with 2D images and hand gestures, in 3D space, it can also be possible to work with 3D hand gestures if depth information is available.

\ifCLASSOPTIONcaptionsoff
  \newpage
\fi




\bibliographystyle{IEEEtran}
\bibliography{Bib}

\begin{thebibliography}{10}
\providecommand{\url}[1]{#1}
\csname url@samestyle\endcsname
\providecommand{\newblock}{\relax}
\providecommand{\bibinfo}[2]{#2}
\providecommand{\BIBentrySTDinterwordspacing}{\spaceskip=0pt\relax}
\providecommand{\BIBentryALTinterwordstretchfactor}{4}
\providecommand{\BIBentryALTinterwordspacing}{\spaceskip=\fontdimen2\font plus
\BIBentryALTinterwordstretchfactor\fontdimen3\font minus
  \fontdimen4\font\relax}
\providecommand{\BIBforeignlanguage}[2]{{%
\expandafter\ifx\csname l@#1\endcsname\relax
\typeout{** WARNING: IEEEtran.bst: No hyphenation pattern has been}%
\typeout{** loaded for the language `#1'. Using the pattern for}%
\typeout{** the default language instead.}%
\else
\language=\csname l@#1\endcsname
\fi
#2}}
\providecommand{\BIBdecl}{\relax}
\BIBdecl

\bibitem{hsi1}
X.~Zhu, J.~Yang, and A.~Waibel, ``Segmenting hands of arbitrary color,''
  \emph{in the Proceedings of the Fourth IEEE, Grenoble , France}, 2000.

\bibitem{templateMatching2}
J.~P. Lewis, ``Fast template matching,'' \emph{Vision Interface 95, Canadian
  Image Processing and Pattern Recognition Society, Quebec City, Canada}, May
  15-19, 1995, p. 120-123.

\bibitem{jong}
J.~S. Bae and T.~L. Song, ``Image tracking algorithm using template matching
  and psnf-m,'' \emph{International Journal of Control, Automation, and
  Systems}, vol. vol. 6, no. 3, pp. 413-423, June 2008.

\bibitem{vina}
V.~M. Lomte, R.~M. Lomte, D.~Mastud, and S.~Thite, ``Robust barcode recognition
  using template matching,'' \emph{International Journal of Engineering
  Sciences and Emerging Technologies}, vol. Volume 2, Issue 2, pp: 59-65, June
  2012.

\bibitem{kalman1}
R.~E. Kalman, ``A new approach to linear filtering and prediction problems,''
  \emph{Transaction of the ASME-Journal of Basic Engineering}, March 1960.

\bibitem{kalman2}
P.~S. Maybeck, ``Stochastic models, estimation, and control, volume 1,''
  \emph{Academic Press, Inc}, 1979.

\bibitem{kalman3}
\BIBentryALTinterwordspacing
G.~Welch and G.~Welch, ``An introduction to the kalman filter,'' \emph{TR
  95-041, University of North Carolina at Chapel Hill}, 1995. [Online].
  Available: \url{http://www.cs.unc.edu/~welch/kalman/}
\BIBentrySTDinterwordspacing

\bibitem{mulder}
A.~Mulder, ``Hand gestures for hci,'' \emph{Technical Report 96-1, vol. Simon},
  1996.

\bibitem{quek}
F.~Quek, ``Towards a vision based hand gesture interface,'' \emph{in
  Proceedings of Virtual Reality Software and Technology, Singapore}, pp. pp.
  17--31, 1994.

\bibitem{mit_sixth_sense}
P.~Mistry, P.~Maes, and L.~Chang, ``Wuw - wear ur world: a wearable gestural
  interface,'' \emph{In Proceedings of the 27th international Conference
  Extended Abstracts on Human Factors in Computing Systems (Boston, MA, USA,
  April 04- 09, 2009)}.

\bibitem{mistry2}
P.~Mistry and P.~Maes, ``Sixth-sense: a wearable gestural inter-face,''
  \emph{In ACM SIGGRAPH ASIA}, vol. pp. 1-1, 2009.

\bibitem{mistry3}
\BIBentryALTinterwordspacing
P.~Mistry, ``Sixthsense integrating information with the real world,'' 2009.
  [Online]. Available: \url{http://www.pranavmistry.com/projects/sixthsense/}
\BIBentrySTDinterwordspacing

\bibitem{kinect2}
T.~Leyvand, C.~Meekhof, Y.~C. Wei, J.~Sun, and M.~Baining~Guo, ``Kinect
  identity technology and experience,'' \emph{Published by the IEEE Computer
  Society}, 2011.

\bibitem{kinect}
\BIBentryALTinterwordspacing
``Kinect.'' [Online]. Available: \url{http://en.wikipedia.org/wiki/Kinnect}
\BIBentrySTDinterwordspacing

\bibitem{matthias}
M.~Baldauf and P.~Fr?hlich, ``Supporting hand gesture manipulation of projected
  content with mobile phones.''

\bibitem{Betke}
M.~Betke, J.~Gips, and P.~Fleming, ``The camera mouse: Visual tracking of
  body,'' \emph{IEEE TRANSACTIONS ON NEURAL SYSTEMS AND REHABILITATION
  ENGINEERING, VOL 10, NO 1,}, 2012.

\bibitem{camera_mouse2}
J.~Gips, M.~Betke, and P.~Fleming, ``The camera mouse: Preliminary
  investigation of automated visual,'' \emph{RESNA 2000 Conference in Orlando},
  pp. RESNA Press, pp. 98--100., 2000.

\bibitem{wolfgang}
\BIBentryALTinterwordspacing
W.~H?rst and C.~V. Wezel, ``Gesture-based interaction via finger tracking for
  mobile augmented reality,'' \emph{Multimedia Tools and Applications}, vol.
  Volume 62, Issue 1, pp 233-258, January 2013. [Online]. Available:
  \url{http://link.springer.com/article/10.1007/s11042-011-0983-y}
\BIBentrySTDinterwordspacing

\bibitem{Swain}
M.~J. SWAIN and D.~H. BALLARD, ``Color indexing,'' \emph{International journal
  of Computer Vision, 7:1, 11-32}, 1991.

\bibitem{james}
J.~L. Crowley, F.~Berard, and J.~Coutaz, ``Finger tracking as an input device
  for augmented reality,'' \emph{in the proceedings of the International
  Workshop on Face and Gesture Recognition}, June 1995.

\bibitem{robert}
R.~Y. Wang, ``Real time hand tracking as a user input device.'' \emph{ACM
  Symposium on User Interface Software and Technology (UIST), October 19-22,
  2008, Monterey, CA}.

\bibitem{Chawalitsittikul}
P.~Chawalitsittikul and N.~Suvonvorn, ``Real time hand marker tracking as a
  user input device,'' \emph{In Proceedings of ECTI Conference on Application
  Research and Development (ECTI-CARD)}, May 10-12, 2010, Chonburi, Thailand,
  p.427-431.

\bibitem{wang}
R.~Y. Wang and Popovic, ``Real-time hand-tracking with a color glove,''
  \emph{In ACM SIGGRAPH 2009 Papers (New Orleans, Louisiana, August 03 - 07,
  2009). H. Hoppe, Ed. SIGGRAPH '09. ACM, New York, NY, 1-8}.

\bibitem{daniel}
D.~S?kora, D.~Sedl?cek, and K.~Riege, ``Real-time color ball tracking for
  augmented reality,'' \emph{Proceedings of the 14th Eurographics Symposium on
  Virtual Environments, pages 9-16, Eindhoven, The Netherlands, May 2008.}

\bibitem{haiqian}
H.~Yu and M.~Leeser, ``Optimizing data intensive window-based image processing
  on reconfigurable hardware boards,'' \emph{IEEE Workshop on Signal Processing
  Systems Design and Implementation}, 2005.

\bibitem{Manresa}
C.~Manresa, J.~Varona, R.~Mas, and F.~J. Perales, ``Real ?time hand tracking
  and gesture recognition for human-computer interaction,'' 2000.

\bibitem{book1}
K.~N. Plataniotis and A.~N. Venetsanopoulos, \emph{Color Image Processing and
  Applications}.

\bibitem{Gonzalez}
R.~C. Gonzalez and R.~E. Woods, \emph{Digital Image Processing}, 2002.

\bibitem{lam}
S.~K. Lam, C.~Y. Yeong, C.~T. Yew, W.~S. Chai, and S.~A. Suandi, ``A study on
  similarity computations in template matching technique for identity
  verification,'' \emph{(IJCSE) International Journal on Computer Science and
  Engineering}, vol. Vol. 02, No. 08, 2010, 2659-2665.

\bibitem{SSDNCC}
K.~G. Derpanis, ``Relationship between the sum of squared difference (ssd) and
  cross correlation for template matching,'' \emph{York University}, December
  2005.

\bibitem{SSDSAD}
``Fast template matching method based optimized sum of absolute difference
  algorithm for face localization,'' \emph{International Journal of Computer
  Applications (0975 ? 8887)}, vol. Volume 18? No.8, March 2011.

\end{thebibliography}

\end{document}